\documentclass[conference]{IEEEtran}
\IEEEoverridecommandlockouts
	
	\usepackage{pifont,bm,multicol,amsfonts,amsmath,color,amssymb,graphicx, epsfig,cite,psfrag,subfigure,algorithm,enumerate,stfloats,algorithm,algorithmic,epstopdf,balance}
\usepackage{tabularx}
	\usepackage{pifont,bm,multicol,amsfonts,amsmath,color,amssymb,graphicx, epsfig,cite,psfrag,subfigure,algorithm,enumerate,stfloats,algorithm,algorithmic,epstopdf,balance}

\begin{document}
%
\title{ Deep Learning Based Antenna Selection for Channel  Extrapolation in FDD Massive MIMO}
\author{\IEEEauthorblockN{Yindi~Yang{$^1$}, Shun~Zhang{$^1$}, Feifei Gao{$^2$}, Chao Xu{$^3$}, Jianpeng Ma{$^1$}, Octavia A. Dobre{$^4$}}
	\IEEEauthorblockA{$^1$State Key Laboratory of  Integrated Service Networks,  Xidian University, Xi'an, China\\
                      $^2${Department of Automation, Tsinghua University, Beijing, China}\\
                      $^3$School of Information Engineering, Northwest A$\&$F University, Yangling, Shanxi, China\\
                      $^4$Faculty of engineering and applied science, Memorial University, St. John¡¯s, Canada}
		Email: \{ydyangdu@163.com,~zhangshunsdu@xidian.edu.cn,\\~feifeigao@ieee.org,~cxu@nwafu.edu.cn,~jpmaxdu@gmail.com,~odobre@mun.ca\}}	
\maketitle
\vspace{-2mm}
\begin{abstract}
In massive  multiple-input multiple-output (MIMO)  systems,  the large number of antennas would bring a great challenge for the acquisition
of the accurate channel state information, especially in the frequency division duplex mode.
To overcome the bottleneck of the limited number  of radio links in  hybrid beamforming, we utilize the neural networks (NNs) to capture the inherent connection between the uplink and downlink channel data sets
{and extrapolate} the downlink channels from  {a subset of the uplink channel state information.}
{We study the antenna subset selection problem in order to achieve the best channel  extrapolation and decrease the data size of NNs.}
The probabilistic sampling theory is {utilized} to approximate
the discrete antenna selection {as} a continuous and differentiable function, which makes
the back propagation  of the deep learning feasible. Then,
we design  the proper off-line training strategy to optimize both
the antenna selection pattern and the {extrapolation} NNs.
Finally, numerical {results} are presented to verify
the effectiveness of our proposed massive MIMO channel {extrapolation} algorithm.
%
\end{abstract}

\maketitle
\thispagestyle{empty}



\IEEEpeerreviewmaketitle

\section{Introduction}

With the increasing demand for the information transmission rate, massive multiple-input multiple-output (MIMO) system has become a key technology for the next generation of wireless communication \cite{efficiency2}.
The {huge} number of antennas in massive MIMO {brings} {a great challenge for  the base station (BS)} to obtain the accurate channel state information (CSI), especially in frequency division duplex (FDD) mode\cite{CSIFeedback}.
In fact, it is possible to utilize microwave scattering reciprocity {between uplink and downlink} to reduce the overhead of the channel acquisition. {In \cite{reconstruction}, the authors
{utilized
the reciprocity property} and  proposed a closed-loop channel estimation scheme for the hybrid massive  MIMO.}
Yu \textit{et al.} \cite{YUHAN} {designed}
an efficient downlink channel reconstruction scheme for the FDD {massive MIMO} system.
In \cite{2019arXiv190502371L}, {Li \textit{et al.}} utilized {the} expectation-maximization   and optimal Bayesian Kalman filter methods to accurately track the {downlink} channel with partial prior knowledge achieved {from the uplink training.}

{Since deep learning (DL) can effectively dig out
the latent and complex relation among different data sets,
{researchers have attempted to} utilize DL for improving the performance
of the massive MIMO channel estimation.}
In \cite{8322184}, Wen \textit{et al.} {constructed a DL-based scheme}
to realize the downlink CSI sensing
and to enhance the quality of CSI reconstruction at BS. {Alkhateeb \textit{et al.} in \cite{Space}
utilized deep neural networks (DNNs) to
approximate the complex mapping function
among  the channels related with different frequency bands and locations.}
 Yang \textit{et al.} in \cite{8795533} proposed a DL-based
 uplink-to-downlink mapping scheme to infer downlink massive MIMO channels from the uplink ones.
 {Choi} \textit{et al.} \cite{access} {developed a  DL extrapolation technique to
 implement the CSI mapping between the downlink and
 the uplink, where the uplink channel path gains of low dimension were treated as
 the input of the neural networks {(NNs).} }

{As mentioned above, the uplink-to-downlink channel {extrapolation} has been widely examined within the massive MIMO {framework}. However, at millimeter wave band, the hybrid beamforming structure is usually
{adopted} to decrease the hardware cost.
Moreover, with the development of the {extremely large} massive MIMO, the hybrid mode may be
the feasible way to enjoy the high spatial resolution.
{Under this structure, all the} uplink CSI of all antenna elements
{cannot} be acquired at the same time \cite{hybird}.
Even though we can scan all the antenna elements to achieve all CSI, it {would spend}
the time resources.
Intuitively, we can utilize the partial uplink CSI observed at {a few} antennas to extrapolate the full downlink one with the power of NNs.

Obviously, the performance of the downlink channel extrapolation from the partial uplink CSI
is closely related with the antenna selection pattern.
If the channels at different antenna elements are
independent, the uniform selection pattern would be the best choice. {However}, in massive MIMO system, the
distance between antenna elements can be small enough that there exists strong correlation among
the channels. Under this scenario, the uniform selection pattern
may not {be} the optimal scheme.
{So}, how to select the antenna subset for uplink channel estimation is very important.
In the model-based MIMO signal processing frameworks, there
are many effective methods to complete the antenna selection through solving the discrete combinatorial optimization \cite{antennaselect1, antennaselect2, antennaselect3}. {However}, the DL-based channel extrapolation mainly lies in the huge data learning without of the accurate model and cannot directly incorporate the traditional antenna selection schemes \cite{8322184}--\cite{access}. Thus, we should design proper DL-based antenna selection and effectively extrapolate the downlink channels from the partial uplink ones. {In this paper}, we resort to the probabilistic sampling theory and model the discrete antenna selection as a continuous and differentiable function. In such way, we can
design  the proper off-line DL training strategy to optimize both
the antenna selection pattern and the extrapolation {NNs} through efficient back propagation.

\begin{figure}[!t]
	\centering
	\includegraphics[width=3.3in]{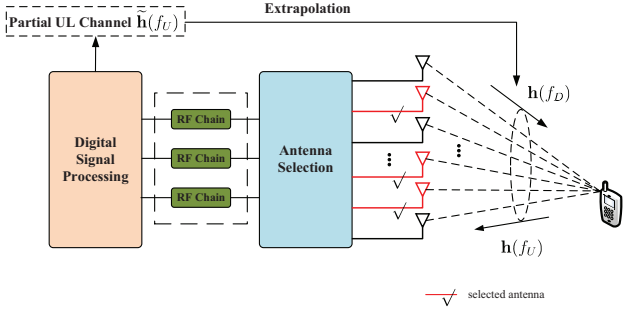}
	\caption{{Our considered problem.}}
	\label{problem}
\end{figure}

\section{Channel Model}


We consider {a}  massive MIMO system, which contains one BS and one user.
{The BS is equipped with $N$ antennas in the
form of {non-uniform linear array (NULA)\footnote{Theoretically, compared with
uniform linear array (ULA), the NULA can obtain a narrower beam without increasing the number of array elements~\cite{NULA}. The irregular physical structure would aggregate the non-uniform distribution of the massive MIMO channel. } \cite{NULA}}, and the user {is} equipped with single antenna.}
{The coordinate vector of the antenna position $\mathbf d$ is
$\mathbf d=[d_1,d_2,\ldots,d_{N}]$. }
Let $\mathbf h(f)$ denote the channel from the user to the BS at frequency $f$. Assume that the channel between the user and the BS consists of $N_p$ paths. Therefore, the  $N\times1$ channel vector $\mathbf h(f)$ can be written as {\cite{deepmimo}}
\begin{align}
\mathbf h(f) = \sum\limits_{i=1}^{N_p}{\alpha_{i}(f)} e^{j\phi_i}e^{-j2\pi f\tau_i}\mathbf{a}({\theta}_{i},f),
\end{align}
{where {$\mathbf h(f)=[h_1(f),h_2(f),\ldots,h_{N}(f)]^T$,} {with {$(\cdot)^{T}$ as the transpose} operator}.
The {$i$-th path} has a delay $\tau_i$, a phase shift $\phi_i$ and a propagation gain $\alpha_i$.} 
Moreover, the spatial steering vector  $\mathbf{a}(\theta_i,f)$ at BS is defined as
\begin{align}
\mathbf{a}(\theta_{i},\!f)\!\!=\!\!\left[\!e^{-j\!\frac{2\!\pi\!fd_1}{c}\!sin\theta_{i}},\! e^{-j\!\frac{2\!\pi\!fd_2}{c}\!sin\theta_{i}}, \!\ldots,\! e^{-j\!\frac{2\pi\!fd_{\!N}}{c}\!sin\theta_{i}}\!\right]^{T},
\end{align}
where $c$ {is the}
speed of light and $\theta_{i}$ denotes the direction of arrival of the {$i$-th path}.


\section{DL-based Antenna Selection}
In FDD, there is a frequency offset between the uplink and downlink channels. Let us denote $\mathbf h(f_U)$ and $\mathbf h(f_D)$ as uplink and downlink channels at frequencies $f_U$ and $f_D$, respectively. Due to the limited number of  radio channels {in the hybrid beamforming structure,}
we assume that not all the elements in
$\mathbf h(f_U)$ can be  achieved, which means that massive MIMO channels are spatially sub-sampled.
In other words, we should determine which antennas should be selected to {extrapolate} the downlink channels.
As shown in Fig.~\ref{problem}, our goal is to utilize these sub-sampled data to {extrapolate} the full elements in $\mathbf h(f_D)$,
where
DNN will be utilized.
In this section, we will successively introduce
the framework design, loss function and learning strategy for the antenna selection based massive MIMO channel {extrapolation.}

\subsection{Framework Design}
As mentioned above, we need to solve the following mapping relationships:
\begin{align}
\mathbf h(f_U) \xrightarrow{\text{sub-sampling}}\widetilde{\mathbf h}(f_U) \xrightarrow{\text{Extrapolation}} \mathbf h(f_D),
\end{align}
where $\widetilde{\mathbf h}(f_U) $ represents the uplink channel after sub-sampling.

\begin{figure}[!t]
	\centering
	\includegraphics[width=3.3in]{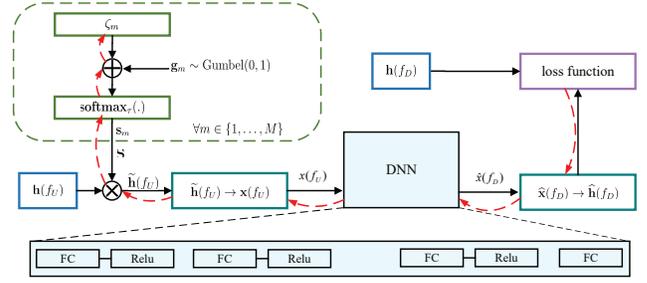}
	\caption{{The architecture of our proposed method}, where the gradient back propagation is shown in red.}
	\label{system architecture}
\end{figure}

{As shown in Fig. \ref{system architecture},} {our framework} contains  two parts, i.e., the antenna selection and the channel {extrapolation}.
 The former sub-samples the original uplink $N\times 1$ vector $\mathbf{h}(f_U)$ to achieve the $M\times 1$ vector
 $\mathbf{\widetilde{h}}(f_U)$ as
\begin{align}
\widetilde{\mathbf{h}}(f_U) = f_{sub}(\mathbf h(f_U)),
\end{align}
where $M<N$ and $f_{sub}(\cdot)$ represents the antenna selection operation.
We can define the spatial compression ratio of the massive MIMO channel as $r=M/N$.
Correspondingly, within the latter part, we resort to the DNN
to reconstruct the downlink channels from
$\widetilde{\mathbf h}(f_U)$ as
\begin{align}
\widehat{\mathbf h}(f_D) = f_{rec}(\widetilde{\mathbf h}(f_U)),
\end{align}
where the $N\times 1$ vector $\widehat{\mathbf h}(f_D)$ represents the recovery version  of $\mathbf h(f_D)$,
and the function $f_{rec}(\cdot)$ is the {extrapolation} operation by DNN.

Our aim is to effectively achieve the antenna selection pattern
and the extrapolation scheme with DL, where the back propagation is usually involved.
However, the antenna selection operation $f_{sub}(\cdot)$ is a discrete combination operation, which hinders
the implementation of the back propagation. To overcome this bottleneck, we introduce
the probabilistic sampling strategy as follows.

\subsection{Learning the Antenna Selection Pattern}
The antenna selection function $f_{sub}(\cdot)$ can be expressed by the $M\times N$ binary sub-sampling matrix
$\mathbf S=[\mathbf s_1^T,\mathbf s_2^T,\ldots,\mathbf s_M^T]^T$\footnote{Here, our aim is to decrease the active antennas at the same time.
Theoretically, different antenna selection patterns can be adopted from time to time.
However, frequently antenna switching may take some signaling overhead for the system and  decrease its energy efficiency. {Accordingly, we consider that the antenna selection operation is stable within a long interval. }},
where the elements of $\mathbf S$ are either 0 or 1,
and the $1\times N$ vector $\mathbf s_m$ contains only one non-zero element. Then, we have
\begin{small}
\begin{align}\label{subsampling}
\widetilde{\mathbf h}(f_U) = \mathbf{S}\mathbf h(f_U).
\end{align}
\end{small}\vspace{-1mm}
Within the probabilistic sampling framework, $\mathbf s_m$ can be defined as {\cite{sub-sampling}}
\begin{align}
\mathbf{s}_m={\mathrm{one\_hot}(z_m)},\label{eqn:s_z}
\end{align}
where $z_m$ is a categorical distributed random variable with the class probabilities $\pi_{m,1},\pi_{m,2},\ldots,\pi_{m,N}$.
Notice that the result of $\mathrm{one\_hot}(z_m)$ is one $N\times 1$  real unit-vector and has only one  non-zero entry, whose index corresponds to the class of the drawn sample.
A larger $\pi_{m,n}$ means that the $m$-th element of $\widetilde{\mathbf h}_U(n)$ would be achieved
from the $n$-th antenna with {higher} probability.
{Different categorical variables, i.e., $z_{m_1}$ and $z_{m_2}$, are independent, $m_1\neq m_2$.}
 Furthermore, we can reparameterize $\pi_{m,n}$ with the unnormalized
log-probabilities $\zeta_{m,n}$ as
\begin{align}
\pi_{m,n}=\frac{\exp(\zeta_{m,n})}{\sum\limits_{n'=1}^{N}\exp (\zeta_{m,n'})}.
\end{align}
\vspace{-1mm}
Here we define the $N\times 1$ vectors $\boldsymbol\pi_{m}=[\pi_{m,1},\ldots,\pi_{m,N}]^T$
and  $\boldsymbol\zeta_{m}=[\zeta_{m,1},\ldots,\zeta_{m,N}]^T$ for further use.
In order to achieve one effective sample from the categorical distribution,
we resort to the $\mathrm{Gumbel}$-$\mathrm{Max}$ trick and generate a realization of $z_m$ as \cite{Gumbel}
\begin{align}
z_m^{\prime}={\arg\max\limits_{n}[g_{m,n}+\zeta_{m,n}]},
\end{align}
where $g_{m,1}$, $g_{m,2}$, \ldots, $g_{m,N}$ are independent and identically distributed (i.i.d.) samples drawn from the
$\text{Gumbel}(0, 1)$ distribution. Correspondingly, $\mathbf s_m$ can be achieved from
$z_m^\prime$ as
\begin{align}
\mathbf s_m=\mathrm{one\_hot}\{\arg\max\limits_{n}[g_{m,n}+\zeta_{m,n}]\}.
\end{align}
However, when we do the above operation from $m=1$ to $m=M$, the same antenna may be repeatly selected.
To avoid this case, we would dynamically exclude {the categories (antennas)}, that have already been chosen,
renormalize the log-probabilities of the rest categories, and then implement
the  $\mathrm{Gumbel}$-$\mathrm{Max}$ trick. Before proceeding, we define $\mathbf g_m=[g_{m,1},g_{m,2},\ldots,g_{m,N}]^T$.

Within {DL} framework, we should iteratively update $\boldsymbol\zeta_m$
through the back propagation
to complete the  antenna selection. However, the operator $\arg\max$ is is not differentiable.
Thus, we will resort to the $\mathrm{softmax}_{\tau}$ function as
a continuous and differentiable approximation of {$\mathrm{one\_hot}\{\arg\max\}$.} Then, we have {\cite{sub-sampling}}
\begin{align}
\mathbf s_m=&\lim_{\tau\rightarrow 0}\mathrm{softmax}_{\tau}(\boldsymbol\zeta_m+\mathbf g_m)\notag\\
=&\lim_{\tau\rightarrow 0}\frac{\exp\{(\boldsymbol\zeta_m+\mathbf g_m)/\tau\}}{\sum\limits_{n=1}^N
\exp\{(\zeta_{m,n}+g_{m,n})/\tau\}},
\end{align}
where the temperature $\tau$ controls the softness of $\mathrm{softmax}_{\tau}$. The lower $\tau$ is, the closer the generated Gumbel-Softmax distribution is to the categorical distribution.
 During training, we will gradually reduce the temperature to  approach the true discrete distribution.
Then, the first-order derivative of $\mathbf s_m$ with respect to $\boldsymbol\zeta_{m}$ can be written as
\begin{align}
\frac{\partial\mathbf s_m}{\partial\boldsymbol\zeta_m^T}=
\frac{\partial}{\partial\boldsymbol\zeta_m^T}\mathbb E_{\mathbf g_m}\left[\mathrm{softmax}_{\tau}(\boldsymbol\zeta_m+\mathbf g_m)\right],~\tau>0.
\end{align}

\subsection{DNN-based  Channel Extrapolation}
The channel {extrapolation} is implemented within DNN $f_{re}(\cdot)$.
Firstly, we reshape the raw input data of DNN,  i.e., $\mathbf{\widetilde{ h}}_U(n)$,  as
\begin{align}\label{input}
\mathbf{x}(f_U)=[\Re (\widetilde{\mathbf h}(f_U))^T, \Im (\widetilde{\mathbf h}(f_U))^T]^T
\end{align}
and input $\mathbf{x}(f_U)$ into DNN.  Correspondingly,
the output of DNN is $\mathbf{\widehat{x}}(f_D)$.
The DNN adopts the fully-connected (FC)
NN architecture with $L$ layers,
The output is a cascade of the nonlinear transformation with respect to $\mathbf x(f_U)$, i.e.,
\begin{align}
\widehat{\mathbf{x}}(f_D)=f^{({L}-1)}_{\boldsymbol\omega_{L-1}}(...f^{(2)}_{\boldsymbol\omega_2}(f^{(1)}_{\boldsymbol\omega_1}(\mathbf x(f_U)))),
\end{align}
where $\boldsymbol\omega_{l}$ is the trainable parameter vector of DNN. Then, each layer computation of the DNN can be expressed as
\begin{align}
f^{(l)}_{\boldsymbol\omega_l}(\mathbf x(f_U)) = t^{(l)}(\mathbf{\bm{\omega}}^{(l)}\mathbf{x}(f_U)+\mathbf{b}^{(l)}), 1 \leq l\leq L-1,
\end{align}
where $\bm{\omega}^{(l)}$ is the weight vector associated with the $(l-1)^{th}$
and $(l)^{th}$ layers, while $\mathbf b^{(l)}$ and $t^{(l)}$ are the bias vector and the
activation function of the $l^{th}$ layer, respectively.

Finally, we can obtain the {extrapolated} massive MIMO downlink channel vector $\widehat{\mathbf h}(f_D)$ from
the real data $\widehat{\mathbf x}(f_D)$.

\subsection{Learning Scheme}

Before proceeding, let us define $\bm\zeta=[\bm\zeta_1^T,\bm\zeta_2^T,\ldots,\bm\zeta_{M}^T]^T$
and $\bm\omega=[\bm\omega_{1}^T,\bm\omega_2^T,\ldots,\bm\omega_{L-1}^T]^T$.
During the network learning stage, we train the model parameters $\bm\zeta$ and $\bm \omega$ by minimizing the
mean squared error (MSE) between the  output $\widehat{\mathbf h}(f_D)$ and
the target $\mathbf h(f_D)$.
Without loss of generality, we use the MSE of the channel estimation as the loss function,
which can be written as
\begin{align}
\mathcal{L}=\frac{1}{NM_{tr}}\sum_{\mu=0}^{M_{tr}-1}\left\|\mathbf h^\mu(f_D)-\widehat{\mathbf h}^\mu(f_D)\right\|_2^2,
\end{align}
where $\|\mathbf a\|$ is the {$L_2$-norm} of  vector $\mathbf a$, and $M_{tr}$ is the batch size.
%
%
Besides, we
promote training towards {one-hot} distributions through penalizing
convergence towards high entropy distribution as
\begin{align}
\mathcal{L_S}=-\sum_{m=1}^{M}\sum_{n=1}^{N}\pi_{m,n}\log\pi_{m,n}.
\end{align}\vspace{-1mm}

When the sub-sampling and {extrapolation} parameters are updated jointly, the resultant optimization problem can be written as:
\begin{align}
\left\{\widehat{\bm\omega},\widehat{\bm\zeta}\right\}=\arg\min_{\bm\omega,\bm\zeta}(\mathcal{L}+\rho\mathcal{L}_s),
\end{align}
where the penalty multiplier $\rho$ evaluates the importance of the different penalties.
Here, the adaptive moment estimation (Adam) \cite{ADAM}
optimizer algorithm is adopted to achieve the optimal model
parameters $\bm\zeta$ and $\bm\omega$. Moreover, we  use different $\eta_\zeta$-learning rate and $\eta_\omega$-learning rate update for $\bm \zeta$ and $\bm \omega$ respectively, where $\eta_\zeta>\eta_\omega$.
\begin{algorithm}[h]
\caption{The {learning steps of}  antenna selection pattern and {extrapolation} network}
\begin{algorithmic}[1]
\REQUIRE Training dataset $\mathcal{D}$, {the number} of iterations $N_{iter}$, $\tau_{start}=5,\tau_{end}=0.5$, and {the initialized} trainable parameters $\bm \zeta$ and $\bm\omega$.
\ENSURE Trained logits matrix $\bm \zeta$ and {extrapolation} network parameters $\bm\omega$.
\STATE Compute $\Delta\tau=\frac{\tau_{start}-\tau_{end}}{N_{iter}-1}$
\FOR{$i=1$ to $N_{iter}$ }
\STATE Draw mini-batches $\mathbf h(f_U)$: a random subset of $\mathcal{D}$
\STATE Draw the reconstructed target: $\mathbf h(f_D)$
\STATE Draw Gumbel noise vectors $\mathbf g_m$ for $m\in\{1,...,M\}$
\STATE Compute $\mathbf s_m=\mathrm{one\_hot}\{\arg\max\limits_{n}[g_{m,n}+\zeta_{m,n}]\}$ and $\mathbf S=[\mathbf s_1;...;\mathbf s_M]$
for $m\in\{1,...,M\}$, and dynamically exclude the repeatedly selected antennas.
\STATE Sub-sample the signal as $\widetilde{\mathbf h}(f_U) = \mathbf{S}\mathbf h(f_U)$
\STATE Achieve the input data of DNN as $\mathbf{x}(f_U)=[\Re (\widetilde{\mathbf h}(f_U))^T, \Im (\widetilde{\mathbf h}(f_U))^T]^T
$
\STATE Compute the output of DNN as $\widehat{\mathbf{x}}(f_D)=f^{({L}-1)}_{\boldsymbol\omega_{L-1}}(...f^{(2)}_{\boldsymbol\omega_2}(f^{(1)}_{\boldsymbol\omega_1}(\mathbf x(f_U))))$
\STATE Compute the  loss function as $\mathcal{L}+\rho\mathcal{L}_s$
\STATE Set $\tau=\tau_{start}-(i-1)\Delta \tau$
\STATE Update $\frac{\partial}{\partial\boldsymbol\zeta_m^T}\mathbb E_{\mathbf g_m}\left[\mathrm{softmax}_{\tau}(\boldsymbol\zeta_m+\mathbf g_m)\right],~\tau>0$
\STATE Use Adam optimer to update $\bm\zeta$ and $\bm \omega$
\ENDFOR
\end{algorithmic}
\end{algorithm}

As mentioned above, the temperature parameter $\tau$ should be gradually decreased
to achieve the discrete distribution. Thus, we set the initialization of $\tau$  as 5.0 and
gradually reduce it to 0.5 during training.
To promote  preservation of the original order, all elements $\zeta_{m,n}$ are initialized as
\begin{align}
\zeta_{m,n}=\beta(n-\frac{N}{M}m)^2+\gamma_{m,n},
\end{align}
where $\beta=-2.73e-3$, $\gamma\sim \mathcal{N}(0,0.01)$,  $m=1,2,\ldots,M$ and $n=1,2,\ldots,N$.

For clarity, we present the detailed learning steps for both
antenna selection and {channel extrapolation} in Algorithm 1.

\section{{Simulation Results}}
In this section, we numerically evaluate the performance of
our proposed {DL and antenna selection based massive MIMO channel {extrapolation} method.}
We first describe the communication scenario and dataset source, and then  introduce the NN parameters. Finally, the performance evaluation of the simulation results is explained. {Moreover, the performance of the DL and uniform antenna selection based channel {extrapolation} is also examined for comparison.}

We consider the indoor distributed massive MIMO scenario `I1'  {of} the DeepMIMO dataset \cite{deepmimo},
which is generated based on {the Wireless InSite software.}
{Correspondingly},  the primary parameters for this case are listed in TABLE~I.
{
For the spacing setting of NULA,
we repeat the vector $0.2[2/3, 6/5, 11/7, 1/8, 4/9, 10/11, 5/12, 3/13, 17/15, 3/16,\\
 1/18, 7/20, 5/21, 1/22, 4/25](\lambda_U)$ of the 16 antennas
four times to achieve the spacing vector for the 64 antennas \cite{Wei2015Peak},
{where
$\lambda_U$ represents the carrier wavelength along the uplink. With respect to ULA, the antenna spacing is set as $0.5\lambda_{U}$}.
Furthermore, within the DeepMIMO dataset, we activate the users located within the region
formed by the 1-st row to the 512-th row.
Then, the number of active users from 1 to 512 is 90862.
The bandwidth of orthogonal frequency division multiplexing (OFDM) is set as 20 MHz, while the number of sub-carriers is 64.
The generated channel samples with  the above parameters are divided into training and testing sets according to the ratio of $4:1$. These data sets are used for the DNN learning and performance evaluation.

\renewcommand\arraystretch{1}
\begin{table}[!t]
\caption{The adopted DeepMIMO dataset parameters. }
\centering
 \setlength{\tabcolsep}{0.5mm}{
\begin{tabular}{|c|c|}
\cline{1-2}
{Parameter}     &{Value}\\
\cline{1-2}
{Name of scenario}     &{I1}\\
\cline{1-2}
{The carrier frequency of uplink and downlink}        &  {2.4GHz, 2.5GHz}\\
\cline{1-2}
{Number of BS antennas in (x, y, z)}        &  {(1, 1, 64)}\\
\cline{1-2}
{Number of paths $N_p$}        &  {5}\\
\cline{1-2}
{Active users}        &  {Row 1 to 502}\\
\cline{1-2}
{System bandwidth}        &  {20 MHz}\\
\cline{1-2}
{Number of OFDM sub-carriers}        &  {64}\\
\cline{1-2}
\end{tabular}}
\end{table}

Each NN layer contains FC and activation function.
 In the hidden layers, the number of neurons is set as $(1024, 1024, 2048, 1024, 512)$ by trails and adjustments, and  $\mathrm{Relu}$  is adopted as the activation function, i.e.,
 $\mathrm{Relu}(x) = \max(x, 0)$. With respect to the input and output layers,
the numbers  of the neurons are same with the sizes of the input and output data vectors, i.e., $\mathbf{x}(f_U)$ and $\widehat{\mathbf{x}}(f_D)$, respectively.
The initial parameters for the learning rate are  $\eta_{\zeta}=0.0005$ and $\eta_{\omega}=0.0001$,
the penalty multiplier $\rho$ is taken as $10^{-8}$, and the batch size is 32.


Fig. \ref{ULA} depicts the channel {extrapolation}  MSE of the proposed method versus the spatial compression ratio $r$. In the figure,
the curves labeled by `uniform'
correspond to  the DL and uniform antenna selection based method, while
the ones marked by `proposed method' represent the performance of our proposed method.
It can be checked that our method can always achieve better {extrapolation}
performance than
the uniform antenna selection based scheme for both ULA and NULA. Specially, compared to the case with ULA, our  scheme can achieve higher performance gain under the NULA scenario, which is because
that the channels from NULA possess  much more non-uniform data structure than that from ULA.
On the other hand, all the MSE curves decrease when $r$ increases from 1/16 to 1/2.

Table II and Table III present the  sequence number of the selected antennas
under different $r$ in NULA and ULA, respectively.
\begin{figure}[!t]
	\centering
	\includegraphics[width=3.5in]{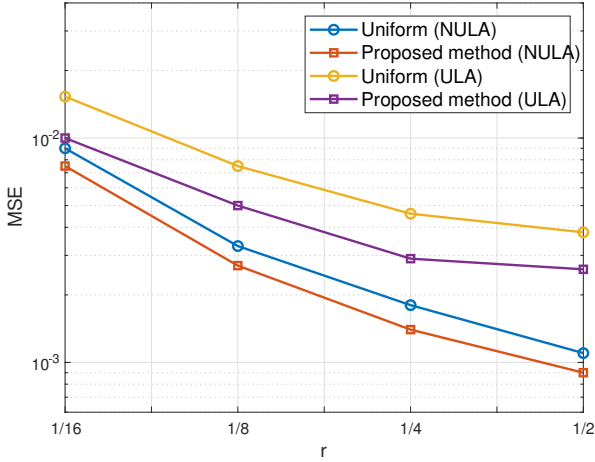}
	\caption{{The MSEs of the downlink channel {extrapolation} versus  the spatial compression ratio $r$}.}
	\label{ULA}
\end{figure}
\renewcommand\arraystretch{1}
\begin{table}[!t]
\caption{The sampled antennas in NULA. }
\centering
 \setlength{\tabcolsep}{0.4mm}{
\begin{tabular}{|c|c|}
\cline{1-2}
{$r$}     &{The sampled antennas}\\
\cline{1-2}
{1/2}     &{1, 5, 6, 7, 9, 10, 13, 14, 16, 17, 20, 22, 25, 26, 29, 31, 33,} \\
\cline{2-2}
{}          &{34, 37, 38, 41, 43, 44, 46, 47, 48, 53, 56, 58, 59, 60, 63}\\
\cline{1-2}
{1/4}        &  {5, 7, 14, 18, 20, 23, 26, 31, 34, 38, 43, 46, 50, 56, 59, 62}\\
\cline{1-2}
{1/8}        &  {6, 14, 23, 31, 38, 46, 54, 62}\\
\cline{1-2}
{1/16}        &  {14, 30, 46, 62}\\
\cline{1-2}
\end{tabular}}
\end{table}
\renewcommand\arraystretch{1}
\begin{table}[!t]
\caption{The sampled antennas in ULA. }
\centering
 \setlength{\tabcolsep}{0.5mm}{
\begin{tabular}{|c|c|}
\cline{1-2}
{$r$}     &{The sampled antennas}\\
\cline{1-2}
{1/2}     &{2, 4, 5, 7, 12, 13, 15, 17, 19, 21, 23, 25, 27, 29, 32, 33, 35,} \\
\cline{2-2}
{}          &{37, 39, 40, 42, 43, 45, 46, 48, 51, 53, 55, 57, 59, 61, 63}\\
\cline{1-2}
{1/4}        &  {3, 7, 11, 16, 19, 23, 27, 30, 35, 39, 43, 47, 51, 55, 59, 62}\\
\cline{1-2}
{1/8}        &  {7, 14, 23, 31, 39, 47, 55, 61}\\
\cline{1-2}
{1/16}        &  {15, 32, 47, 62}\\
\cline{1-2}
\end{tabular}}
\end{table}

Fig. \ref{difference} presents the channel {extrapolation} capability of our method
with different frequency gaps between $\mathbf h(f_U)$ and $\mathbf h(f_D)$, where $r=1/8$.
We consider 4 different frequency gaps for both ULA and NULA.
$\mathbf h(f_U)$ is generated from the first subcarrier at 2.4 GHz  band, while
$\mathbf h(f_D)$ are separately sampled at the
17-th, 33-th, 49-th and 62-th sub-carriers along the 2.5 GHz downlink.
Obviously, as the frequency difference increases,  the MSE increases,
however the performance impact is not very large, which means that the antenna selection method and the NN can achieve good  channel {extrapolation} with big frequency gap.
\begin{figure}[!t]
	\centering
	\includegraphics[width=3.5in]{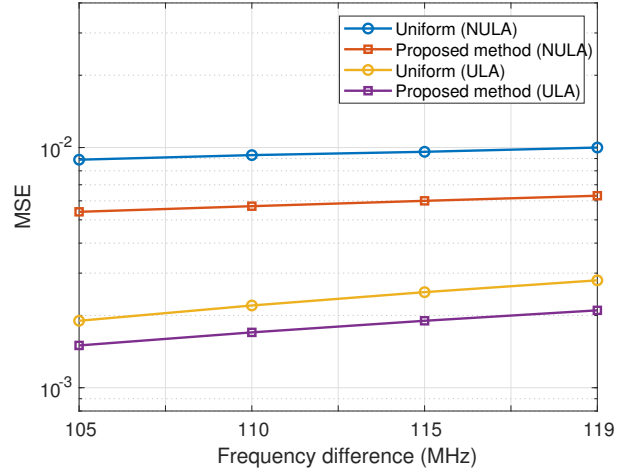}
	\caption{{The  MSEs of the downlink channel {extrapolation} verus  the frequency gaps}.}
	\label{difference}
\end{figure}

\begin{figure}[!t]
	\centering
	\includegraphics[width=3.5in]{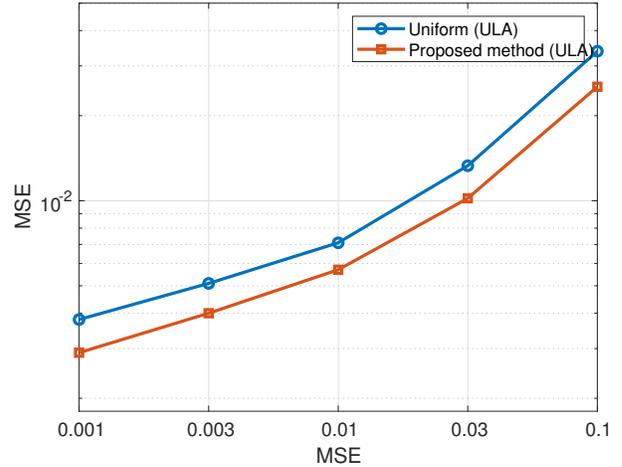}
	\caption{{The channel {extrapolation} MSEs  verus that of the uplink channel estimation}.}
	\label{mse}
\end{figure}

In practice, there exist channel estimation errors along the uplink, which may bring some impacts
on our proposed scheme. Without loss of generality, we model
the uplink channel estimation $\widehat{\mathbf h}(f_U)$ as
$\widehat{\mathbf h}(f_U)={\mathbf h}(f_U)+\mathbf n_{U}$, where
$\mathbf n_{U}$ denotes the additive white Gaussian noise vector.
Then, the variance of $\mathbf n_{U}$ represents
the uplink channel estimation MSE. Correspondingly,
we aim to recover ${\mathbf h}(f_D)$ with
$\widehat{\mathbf h}(f_U)$.
In Fig. \ref{mse}, we evaluate the MSEs of downlink channel {extrapolation}  when the uplink channel estimation
has different MSE values, when ULA is applied. As can be seen from Fig. \ref{mse},
with the decreasing of the uplink channel estimation MSEs, the {extrapolation} performance of the downlink
massive MIMO channels improves within both uniform antenna selection based and our proposed schemes. The uplink channel estimation errors do not affect the performance gain of our scheme over
the uniform antenna selection based one. The two schemes can effectively de-noise the uplink estimated channels under  $\widehat{\mathbf h}(f_U)$'s  high MSE region. The above observation is reasonable
and can be explained as follows. The performance of  DL based {extrapolation} scheme
is determined by both the initial input and the
performance gain of the NN. Moreover, compared with uniform antenna selection based framework,
our scheme utilizes DL to learn better selection pattern.

\section{Conclusion}
In this paper, we examined  DL and antenna selection based massive MIMO channel {extrapolation} scheme.
First, we introduced the probabilistic sampling method to implement the  antenna selection. Then, we inputed the sub-sampled uplink massive MIMO channels into a DNN, and {extrapolated} the full downlink massive MIMO channels with partial uplink CSI.
The Simulation results showed that our proposed scheme  could achieve better performance than the DL and uniform antenna selection based one and
could work effectively with big frequency gaps and uplink channel estimation errors.


\begin{thebibliography}{10}
\providecommand{\url}[1]{#1}
\csname url@samestyle\endcsname
\providecommand{\newblock}{\relax}
\providecommand{\bibinfo}[2]{#2}
\providecommand{\BIBentrySTDinterwordspacing}{\spaceskip=0pt\relax}
\providecommand{\BIBentryALTinterwordstretchfactor}{4}
\providecommand{\BIBentryALTinterwordspacing}{\spaceskip=\fontdimen2\font plus
\BIBentryALTinterwordstretchfactor\fontdimen3\font minus
  \fontdimen4\font\relax}
\providecommand{\BIBforeignlanguage}[2]{{%
\expandafter\ifx\csname l@#1\endcsname\relax
\typeout{** WARNING: IEEEtran.bst: No hyphenation pattern has been}%
\typeout{** loaded for the language `#1'. Using the pattern for}%
\typeout{** the default language instead.}%
\else
\language=\csname l@#1\endcsname
\fi
#2}}
\providecommand{\BIBdecl}{\relax}
\BIBdecl
\bibitem{efficiency2}F.~Rusek, D.~Persson, B.~K. Lau, E.~G. Larsson, T.~L. Marzetta, O.~Edfors, and
  F.~Tufvesson, ``Scaling up {MIMO}: Opportunities and challenges with very
  large arrays,'' {\emph{IEEE Signal Process. Mag.}}, vol.~30, no.~1, pp. 40--60, Jan. 2013.

\bibitem{CSIFeedback}
{S.~Noh, M.~D.~Zoltowski, and D.~J.~Love, ``Training sequence design for feedback assisted hybrid beamforming in massive
MIMO systems," \emph{IEEE Trans. Commun.}, vol. 64, no. 1, pp. 187--200, Jan. 2016.}
%

\bibitem{reconstruction}
{A.~Liao, Z.~Gao, H.~Wang, S.~Chen, M.~Alouini and H.~Yin, ``Closed-loop sparse channel estimation for wideband millimeter-wave full-dimensional MIMO systems," \emph{IEEE Trans. Commun.}, vol.~67, no.~12, pp.~8329-8345, Dec.~2019.}

\bibitem{YUHAN}
{Y. Han, T. Hsu, C. Wen, K. Wong and S. Jin, ``Efficient downlink channel reconstruction for FDD multi-antenna systems," \emph{IEEE Trans. Wireless. Commun.}, vol.~18, no.~6, pp.~3161--3176, Jun.~2019.}

\bibitem{2019arXiv190502371L}
M.~{Li}, S.~{Zhang}, N.~{Zhao}, W.~{Zhang}, and X.~{Wang}, ``{Time-varying
  massive MIMO channel estimation: Capturing, reconstruction and
  restoration},'' \emph{IEEE Trans. Commun.}, vol.~67, no.~11, pp.
  7558--7572, Nov. 2019.










\bibitem{8322184}
C.~{Wen}, W.~{Shih}, and S.~{Jin}, ``Deep learning for massive MIMO CSI
  feedback,'' \emph{IEEE Wireless Commun. Lett.}, vol.~7, no.~5, pp.
  748--751, Oct. 2018.

\bibitem{Space}
{M. Alrabeiah and A. Alkhateeb, ``Deep learning for TDD and FDD
massive MIMO: Mapping channels in space and frequency," in \emph{Proc. 53rd Asilomar Conference on Signals, Systems, and Computers}, Pacific Grove, CA, USA, Nov. 2019, pp. 1465-1470.}

\bibitem{8795533}
Y.~{Yang}, F.~{Gao}, G.~Y. {Li}, and M.~{Jian}, ``Deep learning-based downlink
  channel prediction for FDD massive MIMO system,'' \emph{IEEE Commun.
  Lett.}, vol.~23, no.~11, pp. 1994--1998, Nov. 2019.



\bibitem{access}{
H. Choi and J. Choi, ``Downlink extrapolation for FDD multiple antenna systems through neural network using extracted uplink path gains," \emph{IEEE Access},  vol.~8, pp.~67100--67111, Apr. 2020.
}

\bibitem{hybird}{
O. E. Ayach, S. Rajagopal, S. Abu-Surra, Z. Pi, and R. W. Heath, ``Spatially sparse precoding in millimeter wave MIMO systems," \emph{IEEE Trans. Wireless. Commun.}, vol.~13, no.~3, pp.~1499--1513, Mar. 2014.}

\bibitem{antennaselect1}
Y.~Gao, H.~Vinck, and T.~Kaiser, ``Massive MIMO antenna selection: Switching architectures, capacity bounds, and optimal antenna selection algorithms,'' {\emph{IEEE Trans. Signal Process.}}, vol.~66, no.~5, pp. 1346--1360, Mar. 2018.

\bibitem{antennaselect2}
S.~Asaad, A.~M.~Rabiei, and R.~R.~M\"{u}ller, ``Massive MIMO with antenna selection: Fundamental limits and applications,'' {\emph{IEEE Trans. Wireless Commun.}}, vol.~17, no.~12, pp. 8502--8516, Dec. 2018.

\bibitem{antennaselect3}
P.~V.~Amadori, and C.~Masouros, ``Interference-driven antenna selection for massive multiuser MIMO,'' {\emph{IEEE Trans. Veh. Technol.}}, vol.~65, no.~8, pp. 5944--5958, Aug. 2016.

\bibitem{NULA}
{W.~{Liu} and Z.~{Wang}}, ``Non-uniform full-dimension MIMO: New topologies and opportunities," \emph{IEEE Wireless. Commun.}, vol.~26, no.~2, pp. 124-132, Apr. 2019.

\bibitem{deepmimo}{A. Alkhateeb, ``DeepMIMO: A generic deep learning dataset for millimeter wave and massive MIMO applications," in \emph{Proc. Information
Theory and Applications Workshop (ITA)}, San Diego, CA, Feb. 2019,
pp. 1--8.}

\bibitem{sub-sampling}{
{Iris A.M. Huijben, Bastiaan S. Veeling and Ruud J.G. van Sloun,} ``Deep probabilistic subsampling for task-adaptive compressed sensing," in \emph{Proc. International Conference on Learning Representations},  Addis Ababa, Ethiopia, Apr. 2020.
}
\balance

\bibitem{Gumbel}
E. Gumbel. Statistical theory of extreme values and some practical applications. \emph{NBS
Applied Mathematics Series,} 33, 1954.

\bibitem{ADAM}
{D.~P.~Kingma and J.~Ba, ``ADAM: A method for stochastic optimization,''~\emph{arXiv:1412.6980}, 2014, [Online].~Available: https://arxiv.org/abs/1412.6980}




\bibitem{Wei2015Peak}{
L. Wei, W. Shao, W. Qi, and J. Chen, ``Peak-to-peak search: fast and accurate DOA estimation method for arbitrary non-uniform linear array". \emph{Electron. Lett.}, vol.~51, no.~25, pp. 2078-2080, Dec. 2015.}
\end{thebibliography}
\linespread{1.0}

\end{document}